 \documentclass[preprint2]{aastex}

\newcommand{\RNum}[1]{\uppercase\expandafter{\romannumeral #1\relax}}

\slugcomment{Paper \RNum{2} of \RNum{2}}

\shorttitle{Automatic Detection and Tracking of CMEs \RNum{2}}
\shortauthors{Byrne et al.}

\begin{document}


\title{Automatic Detection and Tracking of CMEs \RNum{2}:\\Multiscale Filtering of Coronagraph Images}


\author{Jason P. Byrne$^1$, Huw Morgan$^{1,2}$, Shadia R. Habbal$^1$ and Peter T. Gallagher$^3$}
\affil{$^1$Institute for Astronomy, University of Hawai'i, 2680 Woodlawn Drive, Honolulu, HI 96822, USA.}
\affil{$^2$Institute of Mathematics and Physics, Aberystwyth University, Ceredigion, Wales, SY23 3BZ.}
\affil{$^3$Astrophysics Research Group, School of Physics, Trinity College Dublin, Dublin 2, Ireland.}
\email{jbyrne@ifa.hawaii.edu\\ApJ, 752, 145. doi:10.1088/0004-637X/752/2/145}



\begin{abstract}
Studying CMEs in coronagraph data can be challenging due to their diffuse structure and transient nature, and user-specific biases may be introduced through visual inspection of the images. The large amount of data available from the SOHO, STEREO, and future coronagraph missions, also makes manual cataloguing of CMEs tedious, and so a robust method of detection and analysis is required. This has led to the development of automated CME detection and cataloguing packages such as CACTus, SEEDS and ARTEMIS. Here we present the development of a new CORIMP (coronal image processing) CME detection and tracking technique that overcomes many of the drawbacks of current catalogues. It works by first employing the dynamic CME separation technique outlined in a companion paper, and then characterising CME structure via a multiscale edge-detection algorithm. The detections are chained through time to determine the CME kinematics and morphological changes as it propagates across the plane-of-sky. The effectiveness of the method is demonstrated by its application to a selection of SOHO/LASCO and STEREO/SECCHI images, as well as to synthetic coronagraph images created from a model corona with a variety of CMEs. The algorithms described in this article are being applied to the whole LASCO and SECCHI datasets, and a catalogue of results will soon be available to the public.
\end{abstract}





\section{Introduction}

Coronal mass ejections (CMEs) are large-scale eruptions of plasma and magnetic field from the Sun into interplanetary space, and have been studied extensively since they were first discovered four decades ago \citep{1972BAAS....4R.394T}. They propagate with velocities ranging from $\sim$\,20~km~s$^{-1}$ to over 2000~km~s$^{-1}$ \citep{2004JGRA..10907105Y, 2001JGR...10629219G}, and with masses of 10$^{14}$\,--\,10$^{17}$~g \citep{1985SoPh..100..563J, 1996ApJ...470..629H, 1992ApJ...390L..37G}, and are a significant driver of space weather in the near-Earth environment and throughout the heliosphere \citep{2010heliophysics, 2005AnGeo..23.1033S}. Traveling through space with average magnetic field strengths of 13~nT \citep{2003SoPh..212..425L} and energies of $\sim$\,10$^{25}$~J \citep{2004JGRA..10910104E}, they can cause geomagnetic storms upon impacting Earth's magnetosphere, possibly damaging satellites, inducing ground currents, and increasing the radiation risk for astronauts \citep{2007A&G....48f..11L}. Thus, models of CMEs and the forces acting on them during their eruption and propagation through the corona remain an active area of research \citep[see reviews by][]{2011LRSP....8....1C, 2012LRSP_Webb}.

The Large Angle Spectrometric Coronagraph suite \citep[LASCO;][]{1995SoPh..162..357B} onboard the Solar and Heliospheric Observatory \citep[SOHO;][]{1995SoPh..162....1D} has observed thousands of CMEs from 1995 to present; and since 2006 the Sun-Earth Connection Coronal and Heliospheric Imaging suite \citep[SECCHI;][]{2008SSRv..136...67H} onboard the Solar Terrestrial Relations Observatory \citep[STEREO;][]{2008SSRv..136....5K} has provided twin-viewpoint observations of the Sun and CMEs from off the Sun-Earth line. Defined as an outwardly moving, bright, white-light feature, CMEs appear in a variety of geometrical shapes and sizes, typically exhibiting a three-part structure of a bright leading front, dark cavity, and bright core \citep{1985JGR....90..275I}. Their geometry is attributed to the underlying magnetic field, generally believed to have a flux-rope configuration. The eruption of the CME is triggered by a loss of stability and its subsequent outward motion is governed by the interplay of magnetic and gas pressure forces in the low plasma-$\beta$ environment of the solar corona. CMEs are commonly linked to filament/prominence eruptions and solar flares \citep{2002ApJ...581..694M, 2002ApJ...566L.117Z}, or labelled `stealth CMEs' if they cannot be associated with any on-disk activity \citep{2009ApJ...701..283R}, but knowledge about their specific driver mechanisms remains elusive. Several theoretical models have been developed in order to describe the forces responsible for the observed characteristics of CME initiation and propagation. The most favoured models explain CMEs in the context of tether straining and release, whereby the outward magnetic pressure increases  due to flux injection or field shearing, and overcomes the magnetic tension of the overlying field \citep{2001AGUGM.125..143K}. Different approaches to such models provide different force-balance interpretations, that lead to a variety of predictions on the kinematic and morphological evolution of CMEs \citep[e.g.][]{2002A&ARv..10..313P, 2003JGRA..108.1410C, 2006PhRvL..96y5002K, 2008ApJ...683.1192L}. To this end, there is a motivation to resolve the observations of CMEs with robust, high-accuracy methods, in order to determine their kinematics and morphology with the greatest possible precision.

From the large number of CMEs observed to date, many exhibit a general multiphased kinematic evolution. This often consists of an initiation phase, an acceleration phase, and a propagation phase which can show positive or negative residual acceleration as the CME speed equalizes to that of the local solar wind \citep{2006ApJ...649.1100Z, 2009SoPh..256..149M}. Statistical analyses can provide a general indication of CME properties \citep[e.g.][]{2000GeoRL..27..145G, 2003AdSpR..32.2637D, 2005AnGeo..23.1033S}, but it remains true that individual CMEs must be studied with rigour in order to satisfactorily derive the kinematics and morphology to be compared with theoretical models. The CME catalogue hosted at the Coordinated Data Analysis Workshop (CDAW\footnote{http://cdaw.gsfc.nasa.gov/CME\_list}) Data Center grew out of a necessity to record a simple but effective description and analysis of each event observed by LASCO \citep{2009EM&P..104..295G}, but its manual operation is both tedious and subject to user biases. Ideally an automated method of CME detection should be applied to the whole LASCO and SECCHI datasets in order to glean the most information possible from the available statistics. A number of catalogues have therefore been developed in an effort to do this, namely the Computer Aided CME Tracking catalogue \citep[CACTus\footnote{http://sidc.oma.be/cactus/};][]{2004A&A...425.1097R}, the Solar Eruptive Event Detection System \citep[SEEDS\footnote{http://spaceweather.gmu.edu/seeds/};][]{2008SoPh..248..485O} and the Automatic Recognition of Transient Events and Marseille Inventory from Synoptic maps \citep[ARTEMIS\footnote{http://www.oamp.fr/lasco/};][]{2009SoPh..257..125B}. However, these automated catalogues have their limitations. For example CACTus imposes a zero acceleration, while SEEDS and ARTEMIS employ only LASCO/C2 data. The motivation thus exists to develop a new automated CME detection catalogue that overcomes such drawbacks, and indeed methods of multiscale analysis have shown excellent promise for achieving this \citep{2009A&A...495..325B}.

In this paper we discuss a new coronal image processing (CORIMP) technique for detecting and tracking CMEs. We outline our application of an automated multiscale filtering technique, to remove small scale noise/features and enhance the larger scale CME in single coronagraph frames. This allows the CME structure to be detected with increased accuracy for deriving the event kinematics and morphology. A companion paper \citep[][hereafter referred to as Paper~\RNum{1}]{2012ApJ...752..144M} outlines the steps used in preprocessing the coronagraph data with a normalizing radial graded filter \citep[NRGF;][]{2006SoPh..236..263M} and deconvolution technique for removing the quiescent background features, leading to a very clean input for the automatic CME detection algorithm. These image processing steps are based on ideas first developed by \citet{2010ApJ...711..631M}, where a more rudimentary approach was taken to isolate the dynamic component of coronagraph images. The new methods of Paper~\RNum{1}, in conjunction with those outlined here, have led to a significant improvement in our ability to automatically detect and track CMEs in coronagraph data, such that a wealth of information on their structure and evolution may be obtained.

In Section~\ref{sect_automation} we outline the multiscale filtering techniques employed, and our method of automatically detecting and tracking CMEs in coronagraph images. In Section~\ref{sect_data} the effectiveness of the CORIMP algorithms is demonstrated through their application to sample cases of LASCO and SECCHI data, and to a model corona with CMEs of known morphology. In Section~\ref{sect_conclusions} we discuss the results and conclusions from the techniques.

\section{Automated CME Detection And Tracking}
\label{sect_automation}

\begin{figure*}[!p]
\centerline{\includegraphics[scale=0.85]{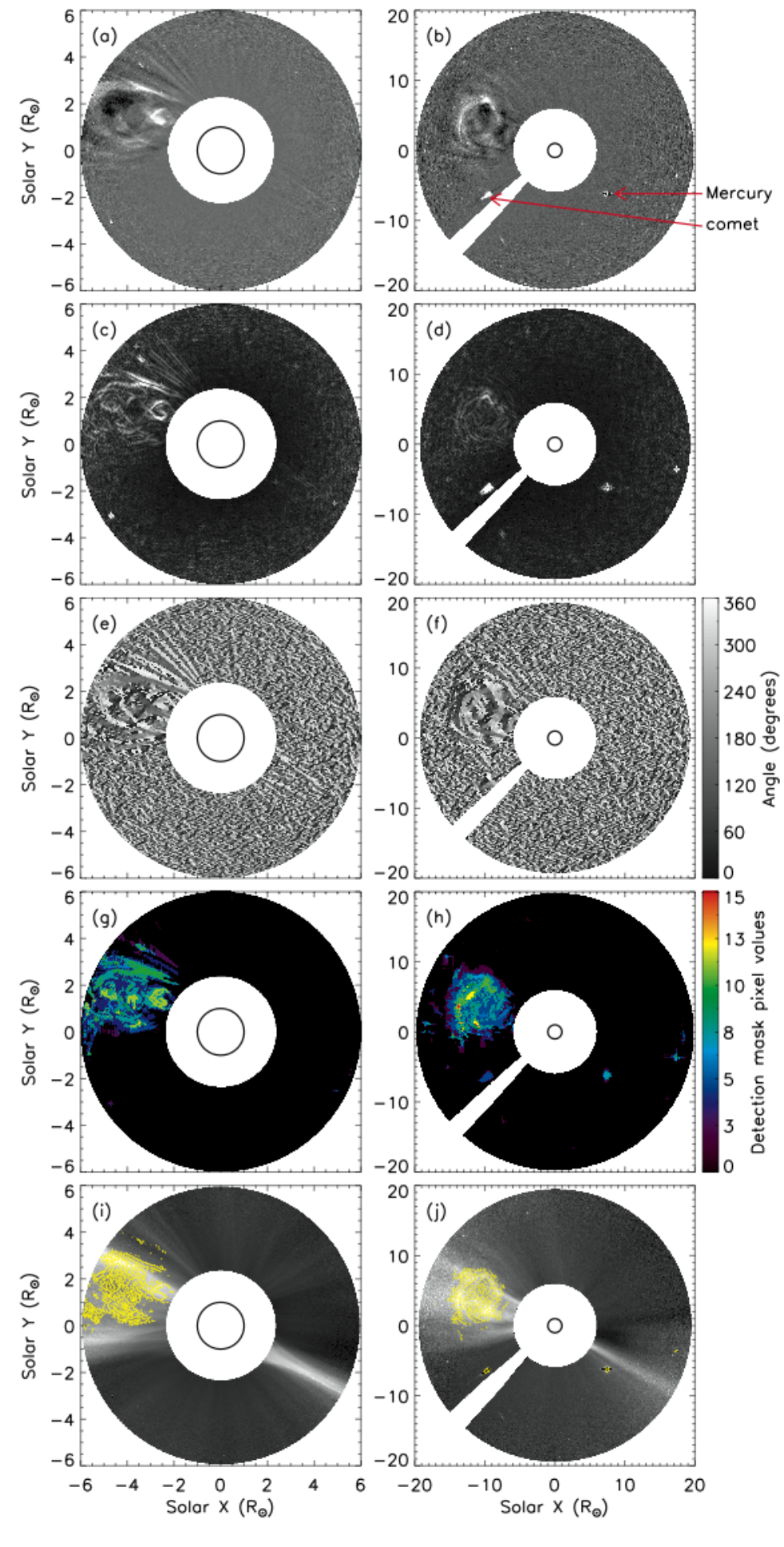}}
\caption{Output of the CORIMP automated techniques applied to images of a CME on 2010 March 12 from the LASCO/C2 (a) and C3 (b) coronagraphs at times 05:06 and 11:42~UT respectively. The images have been processed using the NRGF and quiescent background separation technique outlined in Paper~\RNum{1} to isolate the dynamic coronal structure. A comet and the planet Mercury are also observed in the C3 field-of-view. (c) and (d) show the magnitude information (edge strength), and (e) and (f) show the angular information, at a particular scale of the multiscale decomposition outlined in Section~\ref{sect_multiscale}. (g) and (h) show the resulting CME detection masks following the scoring system outlined in Section~\ref{sect_detectionmask}. (i) and (j) show the final CME structure detection overlaid in yellow on the C2 and C3 images (with reduced intensity scaling to better view the overlaid edges). The edges were determined using a pixel-chaining algorithm on the magnitude and angular information of the multiscale decomposition.}
\label{figure_data10}
\end{figure*}

Figures~\ref{figure_data10}a and \ref{figure_data10}b show a LASCO/C2 and C3 image of a CME on 2010 March 12 at times 05:06 and 11:42~UT respectively, having been processed as per the techniques of Paper~\RNum{1} (namely the NRGF and quiescent background separation). The CME is somewhat ill-defined in the C2 image, consisting of a complex loop structure surrounded by streamer material. In the C3 image the CME has a clearer three-part structure consisting of a bright front, dark cavity, and trailing core. A comet and the planet Mercury are also visible in the C3 field-of-view.   

In order to avoid introducing unwanted edge effects with the automated detection technique, i.e., to prevent the occulter/field-of-view edges from dominating the detection, the image data is numerically reflected inwards and outwards of the occulted field-of-view in the image. This acts to smooth out the sudden change of intensity at the limits of the field-of-view, that would otherwise be detected as significant edges along the boundaries of the image data. The strength of the occulter edges would also suppress the true structure lying close to the occulter edge in the image, somewhat limiting the image area eligible for CME detection by a factor dependent on each particular scale of the multiscale decomposition.

\subsection{Multiscale Filtering}
\label{sect_multiscale}

A multiscale filter is applied as outlined in \citet{2008SoPh..248..457Y} and shown by \citet{2009A&A...495..325B, 2010NatCo...1E..74B, 2011AdSpR..47.2118G, 2011igi-global} to be effective for studying CMEs in coronagraph data. The fundamental idea behind multiscale analysis is to highlight details apparent on different scales within the data. Noise can be effectively suppressed, since it tends to occur only on the smallest scales. Wavelets, as a multiscale tool, have benefits over previous methods, such as Fourier transforms, because they are localised in space and are easily dilated and translated in order to operate on multiple scales. The fundamental equation describing the filter is given by:
\begin{equation}
\psi_{a,b}(t)\,=\, \frac{1}{\sqrt{b}} \, \psi (\frac{t-a}{b})
\end{equation}
where $a$ and $b$ represent the shifting (translation) and scaling (dilation) of the mother wavelet $\psi$ which can take several forms depending on the required use. Here, a method of multiscale decomposition in 2D is employed, through the use of low and high pass filters; using a discrete approximation of a Gaussian, $\theta$, and its derivative, $\psi$, respectively. Since $\theta(x,y)$ is separable, i.e., $\theta(x,y)=\theta(x)\theta(y)$, we can write the wavelets as the first derivative of the smoothing function:
\begin{eqnarray}
\psi_{x}^{s}(x,y)\,&=\, s^{-2} \frac{\partial \theta(s^{-1}x)}{\partial x}\theta(s^{-1}y) \\
\psi_{y}^{s}(x,y)\,&=\, s^{-2} \theta(s^{-1}x)\frac{\partial \theta(s^{-1}y)}{\partial y}
\end{eqnarray}
where $s$ is the dyadic scale factor such that $s=2^j$ for $j=1,2,3,...,J~\in~\mathbb{N}$. Successive convolutions of an image with the filters produce the scales of decomposition, with the high-pass filtering providing the wavelet transform of image $I(x,y)$ in each direction:
\begin{eqnarray}
W_{x}^{s}I \,&\equiv\, W_{x}^s I(x,y)\,=\,\psi_{x}^s (x,y)*I(x,y) \\
W_{y}^{s}I \,&\equiv\, W_{y}^s I(x,y)\,=\,\psi_{y}^s (x,y)*I(x,y)
\end{eqnarray}
Akin to a Canny edge detector, these horizontal and vertical wavelet coefficients are combined to form the gradient space, $\Gamma^s(x,y)$, for each scale: 
\begin{equation}
\Gamma^s (x,y)\, = \,\left[W_{x}^s I,~W_{y}^s I \right]
\end{equation}
The gradient information has an angular component $\alpha$ and a magnitude (edge strength) $M$:
\begin{eqnarray}
\alpha^s(x,y) \, &= \, tan^{-1}\left( W_{y}^s I~/~W_{x}^s I \right) \\
M^s(x,y) \, &= \, \sqrt{ ( W_{x}^s I ) ^2 + ( W_{y}^s I ) ^2 }
\end{eqnarray}
Figures~\ref{figure_data10}c and \ref{figure_data10}d show the magnitude information (with intensity showing the relative edge strengths), and Figures~\ref{figure_data10}e and \ref{figure_data10}f the angular information, for a particular scale ($s=2^{4}$) of the multiscale decomposition applied to the CME images of Figures~\ref{figure_data10}a and \ref{figure_data10}b. As the figures show, the inherent structure of the CME is highlighted very effectively, along with the comet, the planet Mercury, any residual streamer material, and some of the brighter stars. Figures~\ref{figure_data10}e and \ref{figure_data10}f show the angular component $\alpha$ of the gradient, that specifies a direction normal to the intensity regions of the magnitude information $M$. Thus a pixel-chaining algorithm may be employed to trace out all of the multiscale edges in the image, using the orthogonal direction of the angular information as criteria for chaining pixels along the local maxima of the magnitude information.

\subsection{CME Detection Mask}
\label{sect_detectionmask}

The scales upon which the multiscale filtering best resolves the CME have dyadic scale factors of $s=2^{2},\,2^{3},\,2^{4},\,2^{5}$. The discarded finer scales mostly detail the noise, and the coarser scales overly smooth the CME signal. At each of these four scales, the corresponding magnitude $M$ is thresholded at 1.5\,$\sigma$ ($\sigma$ is the standard deviation) above the mean intensity level, resulting from inspection of the method applied to a sample of ten different CMEs of varying speeds, widths and noise levels. This results in regions-of-interest (ROIs) on each image that may be tested as CMEs since they meet the criterion that they are bright features, consequently having stronger edges. To make the 1.5\,$\sigma$ threshold somewhat softer, the initial ROIs are removed and the threshold reapplied at 1.5\,$\sigma$ of the remaining image data to obtain new ROIs. The difference between the new and original ROIs is quantified by subtracting the number of pixels in each, and the intensity threshold reapplied if the subsequent ROI pixel difference is greater than the preceding difference. If the quantified difference decreases, signaling that nothing more can be gained by continuing to soften the threshold on the magnitude image, the threshold is fixed and used to determine the final ROIs. The angular information is then determined for each of these ROIs, since a curvilinear feature will have a wider distribution of angles than a radial feature or a point source in the decomposition. The angular distributions of the individual ROIs are rescaled from ranges 0\,--\,360$^{\circ}$ to 0\,--\,180$^{\circ}$ due to their axial symmetry, and the distribution is normalized to unity. The median value of the distribution across each ROI is then thresholded as a measure for scoring the validity of the detection in order to build up a detection mask of the image:
\begin{enumerate}
\item If the median angular value is $>$\,20\% of the distribution peak then the region is deemed a CME and assigned a score of 3 (the pixels in that ROI are given the value 3).
\item If it is between 10\,--\,20\% the score is 2 (potential CME structure).
\item If it is between 5\,--\,10\% the score is 1 (weak CME structure or part thereof).
\end{enumerate}
Figures~\ref{figure_data10}g and \ref{figure_data10}h show the resulting CME detection mask generated from the additive accumulation of the scores at each scale used for the LASCO/C2 and C3 images. Immediately it is possible to remove the areas of the mask that do not additively achieve a strong enough detection. So, again by inspection across the test sample of ten events, the masks are thresholded at a level $>$\,3 since only the regions that accumulate a sufficient score to be classified as a CME detection are included.

Now that a CME detection has been established, its structure is defined by the edges determined in the pixel-chaining algorithm applied at the scale of $s=2^{4}$, since this scale most consistently exhibits the highest signal-to-noise ratio for the ten sample events. Since the CME detection mask is built with four scales of increasing filter size, there is a possibility that it overshoots the true CME edges in the image and includes unwanted noise surrounding the CME. This effect is somewhat reduced by removing the lowest scoring regions of the detection mask as discussed above, but is further corrected for by eroding the detection mask by a factor of 8 pixels. This factor is half the filter width at scale $s=2^{4}$ chosen based on the fact that if the lowest scale ($s=2^{5}$) ROIs in the detection mask have been removed, then the CME edge being detected will likely be situated half the next filter width ($s=2^{4}$) inside of the detection mask boundary. Figures~\ref{figure_data10}i and \ref{figure_data10}j show the resultant CME structure detections overlaid on the original images. While there is still an element of noise in the detections, clear structure along the twisted magnetic field topology of the erupting CME plasma is defined - and automatically so.

\subsection{Determining The Physical Characteristics Of Detected CMEs}
\label{sect_kins}

\begin{figure*}[!p]
\centerline{\includegraphics[scale=0.35, clip=true, trim=0 120 0 100]{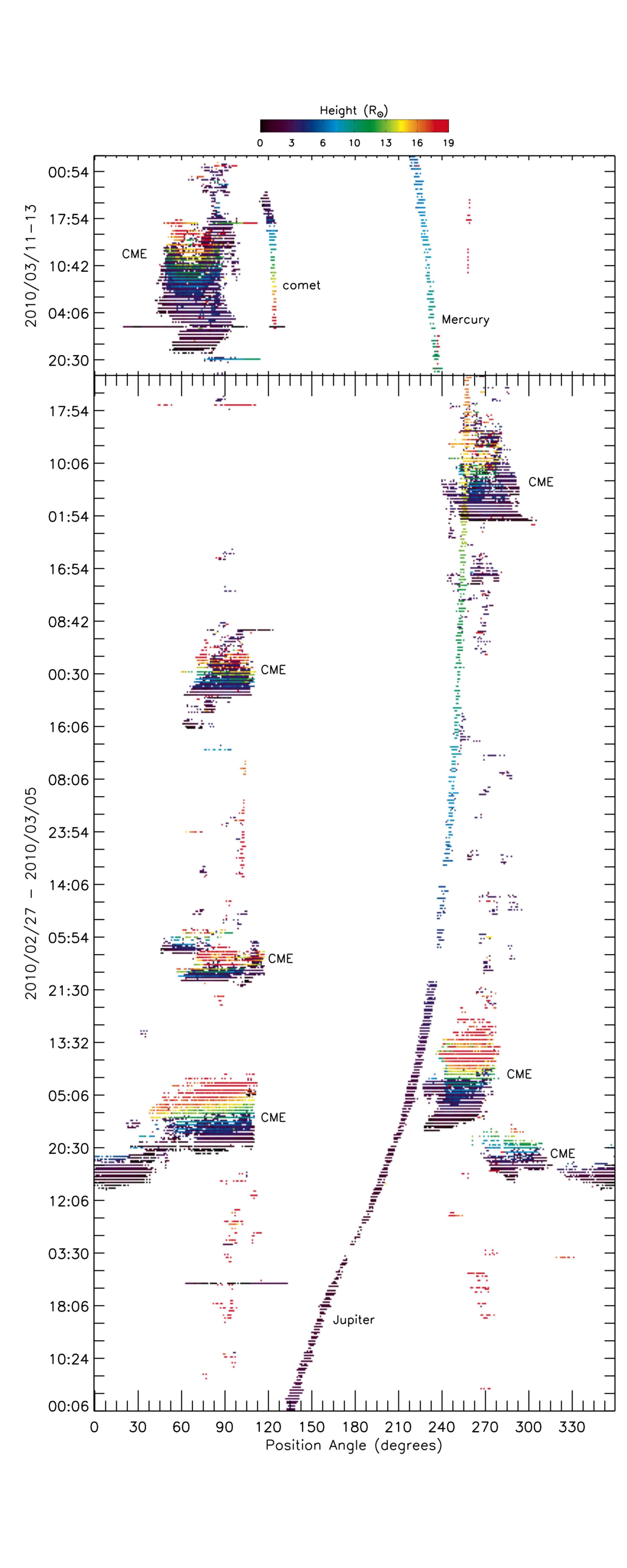}}
\caption{The resulting CME detection stack from the CORIMP automated algorithms applied to the LASCO data for the 2010 March 12 CME shown in Fig.~\ref{figure_data10} (top), and for the interval from 2010 February 27 to March 5 as an example of several typical detections (bottom). The CMEs are clear from their increasing height profiles, versus the decreasing height profile of the comet. The transits of Mercury (top) and Jupiter (bottom) across the images are also clear. Some possible small-scale flows, residual noise and artifacts are also apparent in the detection stack, either as random or somewhat persistent detections of varying color.}
\label{figure_twostack}
\end{figure*}

The outermost points along the strongest detected edges of the CME structure provide the CME height from Sun-center in each image. To determine these so-called strongest edges, the magnitude information deduced from each of the four scales in use here, are multiplied together to enhance the strongest features, and the resulting strengths are assigned to the relative pixels of the edge detections. One median absolute deviation above the median strength of the edges is used as a threshold for determining the strongest edges within the detected CME structure (as opposed to one standard deviation above the mean, which is too easily affected by bright stars, planets, noisy features etc.). The outermost points of these strongest edges, measured along radial lines drawn at 1$^{\circ}$ position angle intervals, are recorded as the span of CME heights in each frame.

As the detections are performed through time, the information from them may be collated into a three-dimensional stack of `Time' versus `Position Angle' versus `Height'. A CME detected at a particular span of position angles through a sequence of frames will appear as a block of variable height in the detection stack, an example of which is shown in Figure~\ref{figure_twostack} (following some further processing via a cleaning algorithm outlined below). The detection stack for the LASCO data containing the CME shown in Figure~\ref{figure_data10} is illustrated in the top part of Figure~\ref{figure_twostack}; while the detection stack for the interval from 2010 February 27 to March 5 is illustrated in the bottom part, as an example of several typical detections during an active period when Jupiter was also in the field-of-view. The angular span of the detections is indicative of the angular width of the CME. Trailing material contained within the internal structure of the CME will also be apparent on the detection stack as the CME front moves out of the field-of-view. Any residual streamer flows that are detected will also appear in the detection stack, though they should only span small angular widths. Because the 2010 March 12 CME has a lot of internal and trailing material, the persistent C2 material detections underly the increasing C3 height detections, as indicated by the somewhat constant purple shade embedded in the CME-specific region of the detection stack in the top of Figure~\ref{figure_twostack}. This example demonstrates how the codes fare with typical issues faced in CME image data, while also demonstrating its success alongside the additional comet and planet detections. Other CMEs will have cleaner profiles than this one, as some of the detections in the bottom plot show. The comet detection height profile shows a decreasing color intensity in time due to its decreasing height as it falls toward the Sun. The planets Mercury and Jupiter show a change in position angle, along with a slight change in height, as they traverse the fields-of-view. Some random detections due either to small-scale flows, noisy features or artifacts in the images are also apparent in the detection stack, mostly concentrated along the streamer belts centered at position angles $\sim$\,90$^{\circ}$ and $\sim$\,270$^{\circ}$.

For the purposes of cataloguing CMEs, a cleaning algorithm was developed and applied to the detection stack to remove much of the noise. The detection stack regions corresponding to CMEs may be automatically isolated by the following criteria:
\begin{enumerate}
\item Detections that lie within two time steps of each other are grouped.
\item Detections that span $<$\,7$^{\circ}$ and do not have adjoining detections within 7$^{\circ}$ are discarded (chosen to match the original threshold in CACTus based on the smallest widths in CDAW. Although since this threshold is implemented after the detections have been made, a lower threshold may be defined for direct comparison with the second version of CACTus if so desired).
\item Detections that have not been grouped with at least two other detections are discarded.
\end{enumerate}
The resulting detection stack provides a cleaner output for determining the CME kinematics and morphology. The height information at each position angle of the isolated detection groups may be recorded and used to build a height-time profile across the angular span of the detection. Since there exists the possibility that persistent C2 detections can underly the C3 detections of a CME, conditions are imposed on the code to retrieve the height-time profile in a manner such that once the CME height along each position angle moves beyond the C2 field-of-view, only its subsequent heights within the C3 field-of-view are recorded. Examples of CME height-time profiles recorded in this way are shown in Section~\ref{subsect_data}. Changes in the angular width of a CME detection may also be recorded as an indicator of its expansion. Thus a final output of information on each CME detection can include CME height, observation time, position angle, trajectory, and angular width. Due to the various methods available for determining kinematics from height-time measurements (e.g., standard numerical differentiation techniques, spline fitting techniques, inversion techniques; see \citealt{2010ApJ...712.1410T} for example), an investigation of the best approach for cataloguing the specific kinematics of CMEs is postponed to future work. Furthermore, the morphological information that can be attained with these methods, arising from the pixel-chained edge detections and overall enhancement of structure within the CME, will facilitate future detailed inspections of the observed ejection material.

\section{Testing On Real And Synthetic Data}
\label{sect_data}

In order to test how well the CME structure is resolved by these automated methods, the algorithm is applied to a selection of real data from the LASCO and SECCHI coronagraphs, and to synthetic data comprising a model corona through which CMEs of various appearance are propagated. We consider first the real data, which is processed according to the methods outlined in Paper~\RNum{1}, namely the NRGF and dynamic separation techniques.

\subsection{LASCO And SECCHI CME Data}
\label{subsect_data}

The automated CME detection technique is applied to LASCO/C2 and C3, and SECCHI/COR2-A and B coronagraph images. For these algorithms, the C2 images have a workable field-of-view of 2.35\,--\,5.95\,R$_{\odot}$, while C3 is limited to 5.95\,--\,19.5\,R$_{\odot}$  (of a potential $\sim$\,30\,R$_{\odot}$) since the signal-to-noise ratio is too low in the outermost portion of its field-of-view to be used for automatically identifying CMEs via these methods. The SECCHI coronagraphs have a workable field-of-view of 3\,--\,11\,R$_{\odot}$ for COR2-A, and 4.5\,--\,11\,R$_{\odot}$ for COR2-B (of a potential $\sim$\,3\,--\,15\,R$_{\odot}$), again limited by the low signal-to-noise ratio. COR1 proved unfeasible for analysis since the non-radial profile of the corona at heights $<$\,3\,R$_\odot$ does not fare well with the NRGF, and the images have too low a signal-to-noise ratio for the automated techniques to operate satisfactorily.

\begin{figure*}[!p]
\centerline{\includegraphics[scale=1, clip=true, trim=0 0 0 0]{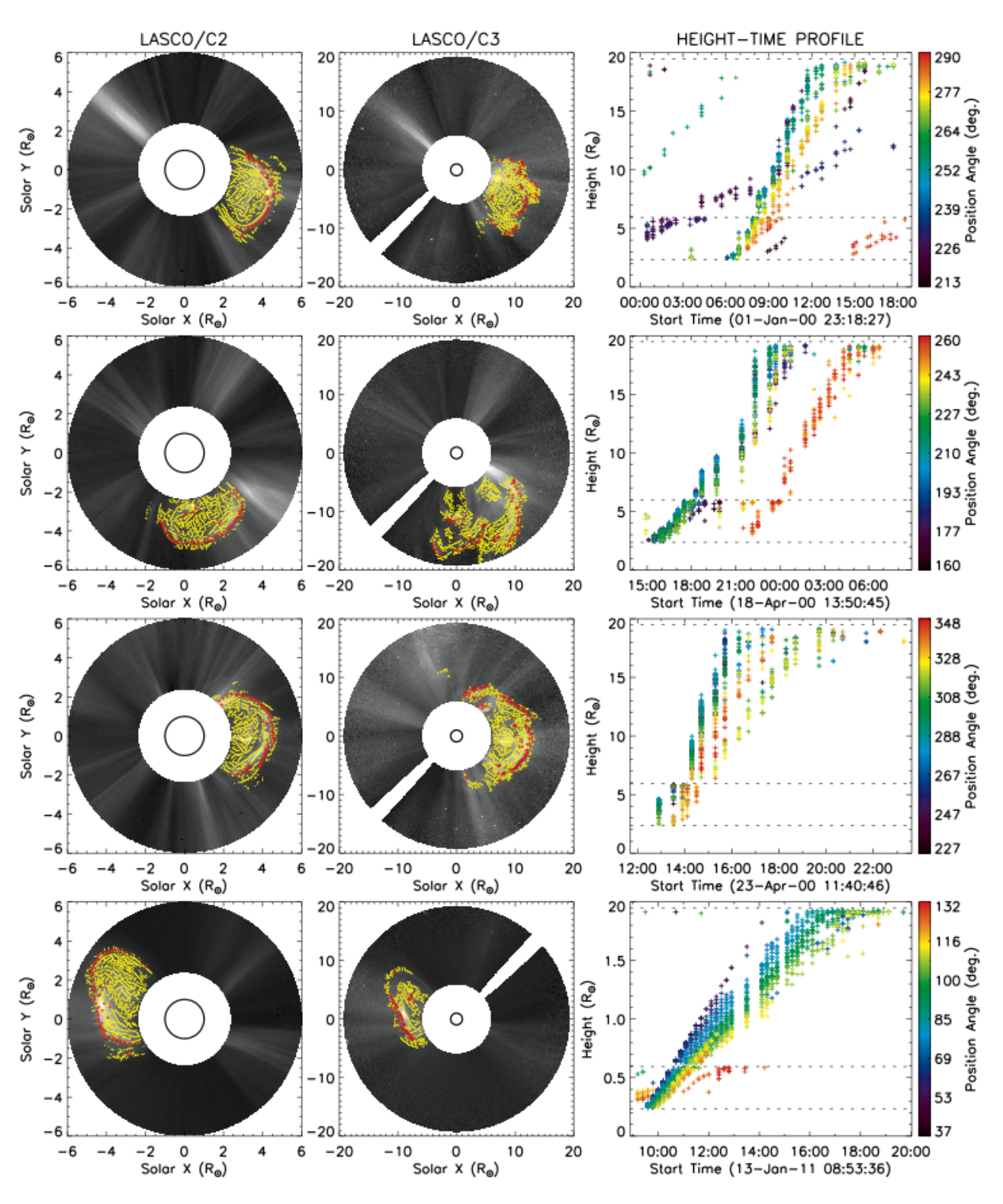}}
\caption{A sample output of the CORIMP automatic CME detection and tracking technique applied to LASCO/C2 and C3 images for 2000 January 2, 2000 April 18, 2000 April 23, and 2011 January 13. Instances of the detections in C2 and C3 are shown for each event, along with the resulting height-time profile corresponding to the tracks of the strongest outermost front (red points on CME) of the overall detected structure (yellow points on CME). Each height-time profile has an associated colorbar that indicates the relevant position angle along which the heights are measured within the angular span of the CME, counter-clockwise from solar north. A corresponding animation of these events is shown in the online material.}
\label{figure_events}
\end{figure*}

\begin{figure*}[!t]
\centerline{\includegraphics[scale=1, clip=true, trim=0 0 0 0]{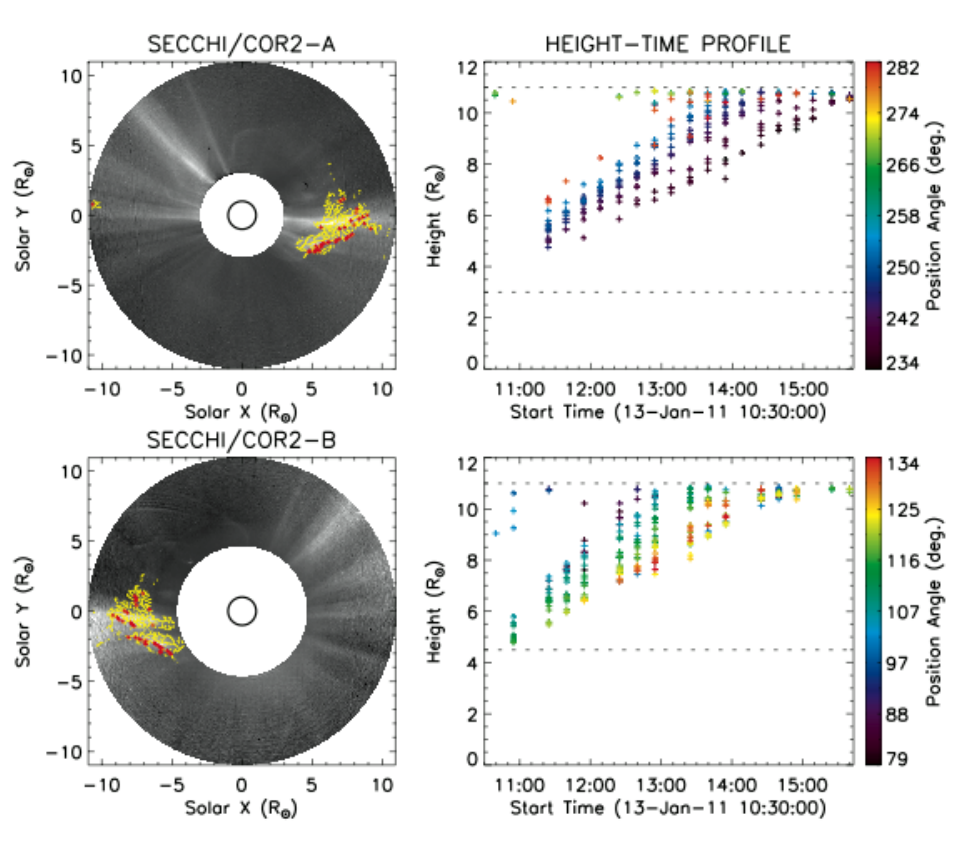}}
\caption{A sample output of the CORIMP automatic CME detection and tracking technique applied to SECCHI/COR2 A and B images for 2011 January 13. The CME appears as a partial halo in the STEREO observations, and parts of its front are too faint to be fully detected in the images. A corresponding animation of this event is shown in the online material.}
\label{figure_events_secchi}
\end{figure*}

Figure~\ref{figure_events} (and its online animation) shows a sample output of the automated detections on LASCO observations of CMEs dated 2000 January 2, 2000 April 18, 2000 April 23, and 2011 January 13. The four CMEs are shown for instances of their detection in C2 and C3, along with the resulting height-time profiles corresponding to the tracks of the strongest outermost front (red points on CMEs) of the overall detected structure (yellow points on CMEs). Each of these profiles has an associated colorbar that indicates the relevant position angle along which the heights are measured. The 2000 January 2 CME exhibits a multi-loop structure, and its height-time profile indicates a relatively constant velocity as it catches up to slower moving material along its southern path. The 2000 April 18 CME has a typical 3-part structure, and its height-time profile shows early acceleration, and some trailing ejecta along its western flank. The 2000 April 23 CME is a highly impulsive partial halo, and its height-time profile is accordingly steep. Finally, the 2011 January 13 CME exhibits asymmetric expansion as the southern portion trails the faster northern front, and its height-time profile thus shows a broadening of speeds across the angular span of the event.

Figure~\ref{figure_events_secchi} (and its online animation) shows the resulting output for the SECCHI/COR2-A and B observations of the 2011 January 13 CME. Note that the CME appears as a halo from the perspective of the STEREO Ahead and Behind spacecrafts. This represents the most difficult class of events to be automatically detected, since halos tend to be faint and somewhat disjoint in the images, sometimes failing to surpass the detection thresholds. Thus, as has happened here, parts of a halo CME can go undetected.

It is at this point that a user may decide how best to treat the CME measurements; for example, by applying a numerical derivative to the height-time measurements to determine velocity and acceleration profiles, or fit a spline of order $k$, say, or any specific model to be tested against the data. In order to test the robustness of the automatically determined CME measurements, model CMEs of known speeds and morphologies are analyzed and the resulting detections inspected in the following section.

\subsection{Model CME Data}

We consider the model data generated from a tomographic reconstruction of the coronal density over a two-week set of observations centered on 2005 January 18 (CR~2025.6) \citep{2009ApJ...690.1119M}. Three model CMEs are generated from a hollow flux-rope connected to the Sun at its footpoints, and another three CMEs are generated from simple plasma blobs of varying density. Observational images of the model data are generated in the likeness of LASCO images, with random Gaussian noise added. The model images are NRGF processed and the dynamic separation technique applied (see Paper~\RNum{1} for details on these models and processing techniques).

\begin{figure}[!t]
\centerline{\includegraphics[scale=0.78, clip=true, trim=0 0 0 0]{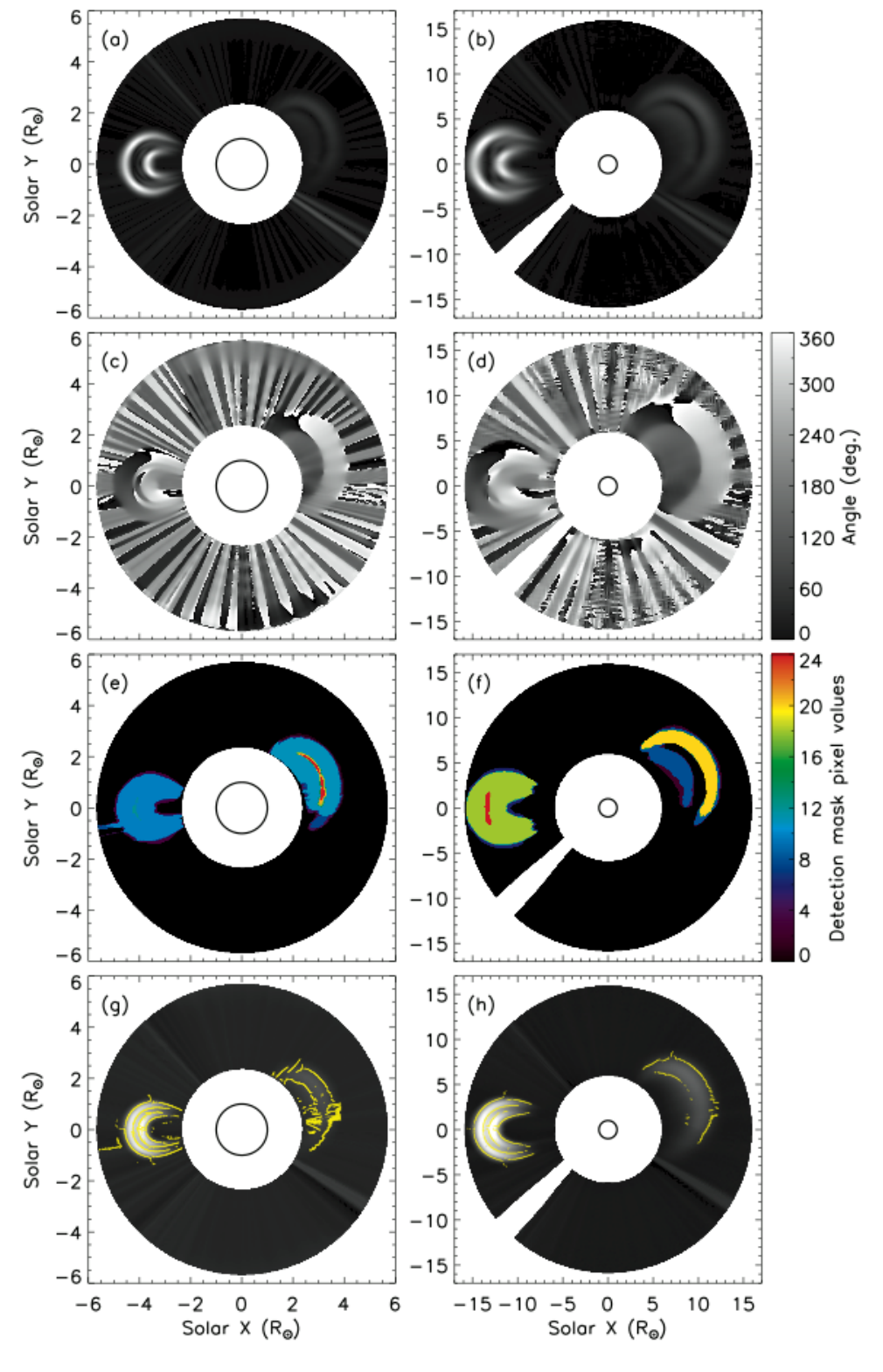}}
\caption{A snapshot of the algorithms applied to the model CMEs A and B. (a) and (b) show the magnitude information (edge strength), and (c) and (d) show the angular information, at a particular scale of the multiscale decomposition outlined in Section~\ref{sect_multiscale}. (e) and (f) show the resulting CME detection masks following the scoring system outlined in Section~\ref{sect_detectionmask}. (g) and (h) show the final CME structure detection overlaid in yellow on the model C2 and C3 images. The edges were determined using a pixel-chaining algorithm on the magnitude and angular information of the multiscale decomposition.}
\label{figure_model8}
\end{figure}

\begin{figure}[!t]
\centerline{\includegraphics[scale=0.78, clip=true, trim=0 0 0 0]{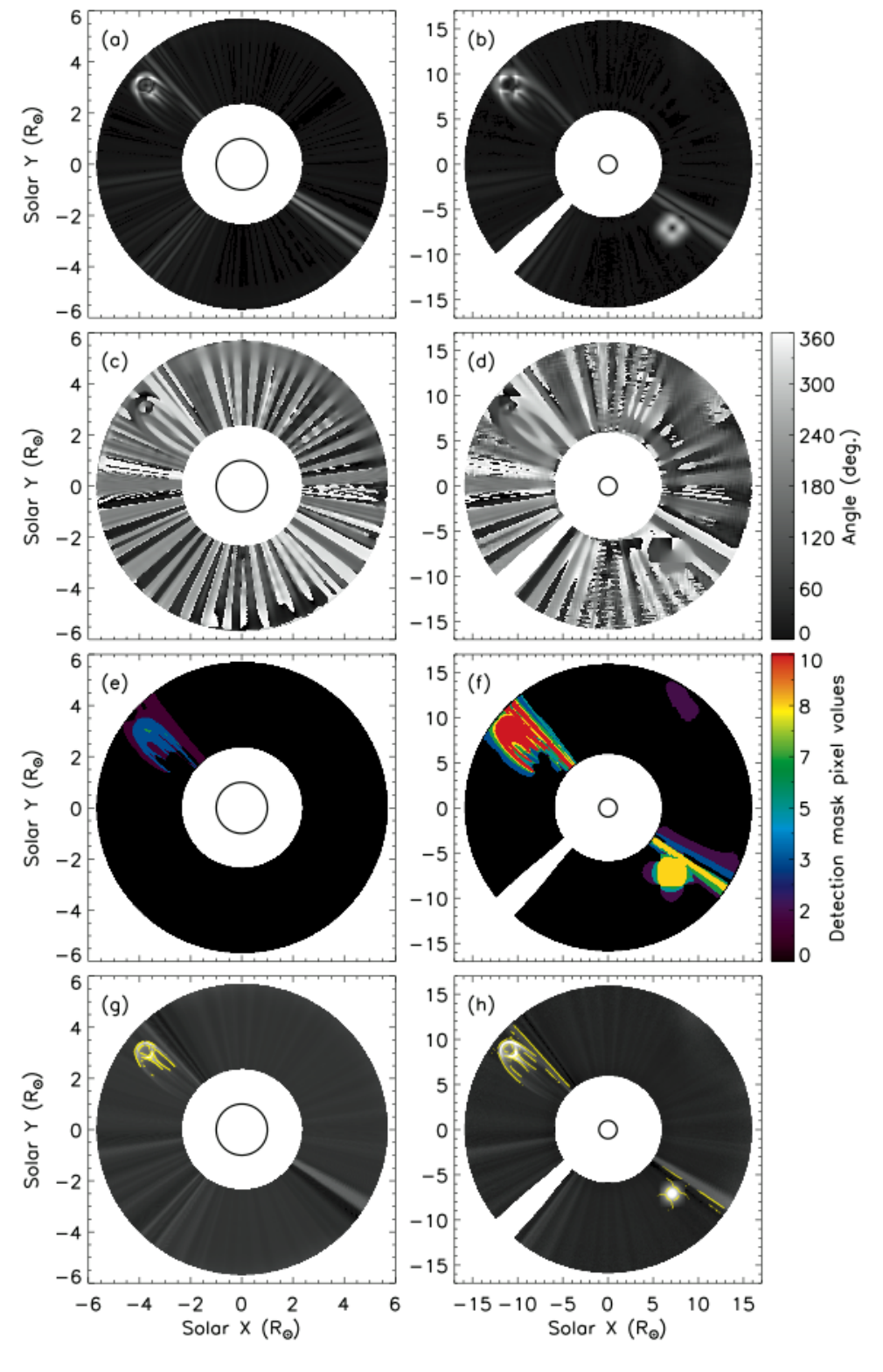}}
\caption{A snapshot of the algorithms applied to the model CMEs C and D (though CME D is only visible here in the C3 image due to its later launch time than CME C). Images displayed as in Figure~\ref{figure_model8}.}
\label{figure_model8_2}
\end{figure}

The CORIMP automated detection and tracking algorithms are applied to the processed model CME data. This allows a qualitative inspection of the resulting edge detections of the model CMEs in the images, with impressive results. Figure~\ref{figure_model8} shows the detections for the case of two CMEs observed simultaneously: CME A launched off the east limb with inclination 90$^{\circ}$ to the observer (edge-on), and CME B launched to the north-west with inclination 70$^{\circ}$ and larger size. It is important to consider that multiple CMEs of different brightness intensities may erupt simultaneously, especially during solar maxima. Specifically halo CMEs (those that propagate toward or away from the observer) tend to be fainter than limb events, due to the Thomson scattering geometry and line-of-sight considerations \citep{2006ApJ...642.1216V, 2009SSRv..147...31H}. This also means that their structure often appears disjointed, which can lead to multiple region detections on a single halo event. This was a strong motivator for dynamically softening the intensity threshold on the magnitude information of the multiscale decomposition, as discussed in Section~\ref{sect_detectionmask}. Figures~\ref{figure_model8}a and \ref{figure_model8}b show the magnitude information, which reveals the residual streamer structure in the radial intensity profile of the model corona, and shows the relative edge strengths of the two flux-rope CMEs as they propagate outward. Figures~\ref{figure_model8}c and \ref{figure_model8}d show the corresponding angular information, conveying the curvilinear nature of the CMEs as compared to the radial structure of the corona. The magnitude and angular information from the optimum four scales of the multiscale decomposition are used to generate the CME detection masks shown in Figures~\ref{figure_model8}e and \ref{figure_model8}f. In these masks the pixel values have been assigned a score corresponding to the strength of the detection (see Section~\ref{sect_detectionmask}). The final edge detections are over-plotted on the original model data in Figures~\ref{figure_model8}g and \ref{figure_model8}h to highlight the structure in the model CMEs.

Figure~\ref{figure_model8_2} is displayed in the same manner as Figure~\ref{figure_model8} for a flux-rope (CME C) launched to the north-east with inclination 50$^{\circ}$, and a density blob (CME D) launched to the south-west alongside a relatively bright streamer region. (The timing of the events is such that CME D is only visible in the C3 image here.) The structure of the bright front of CME C is satisfactorily detected, while its fainter legs are indistinguishable from the background corona. In the C3 images the residual streamer material alongside CME C is included in the detection. The same is true for CME D which is detected along with the residual south-west streamer material. The trailing material from the preceding passage of CME B is also present and detected at its trailing legs in the north-west and beside the top of the residual south-west streamer.

Two final blob CMEs (labelled E and F), with consecutively lower intensities than CME D, are also propagated along the same trajectory as CME D to further test the automated routines. Each of the CME blobs is also satisfactorily detected even at such low intensity levels (note from Paper~\RNum{1} that CME F has a density only 10\% that of streamers at the same height).

In summary, the algorithm is proven to be successful at detecting each of the different model CMEs (a typical limb event, partial halo, narrow flux-rope, and small faint blobs), thus serving as a testament to its effectiveness in creating a real data catalogue. Full halo CMEs represent the limiting case of these events, wherein parts of the faint CME structure may not overcome the thresholds and result in disjoint or incomplete detections.

\subsection{CME Model Kinematics}

\begin{figure}[!t]
\centerline{\includegraphics[scale=0.57, clip=true, trim=0 0 0 0, width=\linewidth]{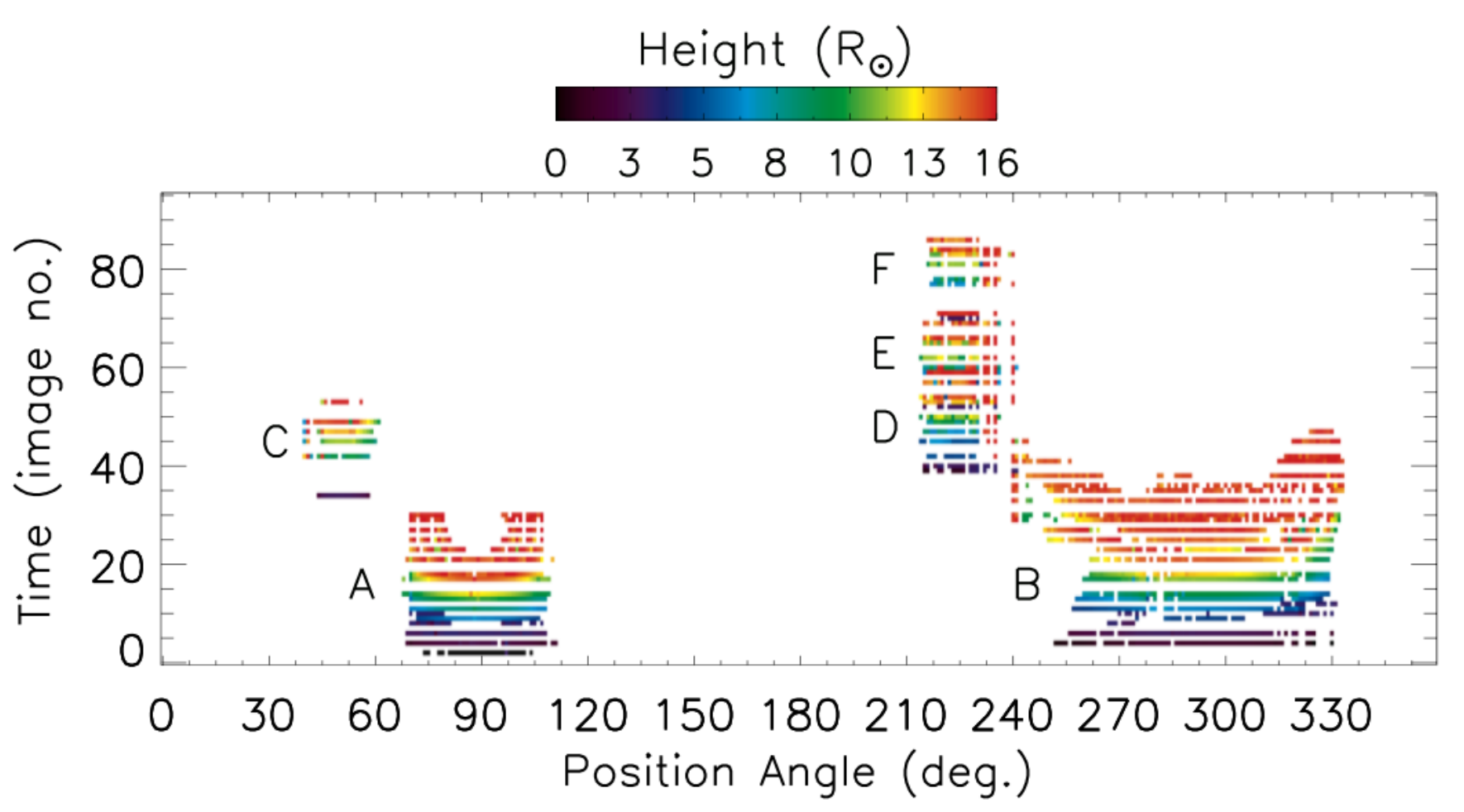}}
\caption{The model CME detection stack, plotted in time, i.e., image number, against position angle measured counter-clockwise from solar north. The intensity corresponds to the height of the outermost points in the detection relative to Sun-center.}
\label{figure_model_stack}
\end{figure}

The model CMEs may be tested for their kinematics by investigating the detection stack that is produced from the automated algorithms. As described in Section~\ref{sect_kins}, the detection stack is generated from the height measurements of the strongest outermost edges (along radial lines drawn from Sun-center) on the detected CME structure at each time step, i.e., for each image. It must be noted that for the model CMEs the median absolute deviation threshold on the strength of the edge detections was not applied since the models are so clean (having very smooth boundaries and minimal internal structure) that this further thresholding is not appropriate for retrieving and testing the model kinematics. It only serves as an additional step to deal with the complexity of edge detections in the real data.

For the presented model CMEs the resulting detection stack is shown in Figure~\ref{figure_model_stack}, with time step plotted against position angle, and intensity representing height from Sun-center. Inspecting Figure~\ref{figure_model_stack} reveals four main detection areas: two distinct regions centered at position angles $\sim$\,90$^{\circ}$ and $\sim$\,50$^{\circ}$ corresponding to CMEs A and C respectively; a large region spanning $\sim$\,250\,--\,340$^{\circ}$ that corresponds to CME B; and a somewhat adjoining region between $\sim$\,215\,--\,240$^{\circ}$ that corresponds to the three density blobs (CMEs D, E, F) that are detected alongside the residual streamer material centered at $\sim$\,240$^{\circ}$. 

\begin{figure}[!t]
\centerline{\includegraphics[scale=0.65, clip=true, trim=0 0 0 0]{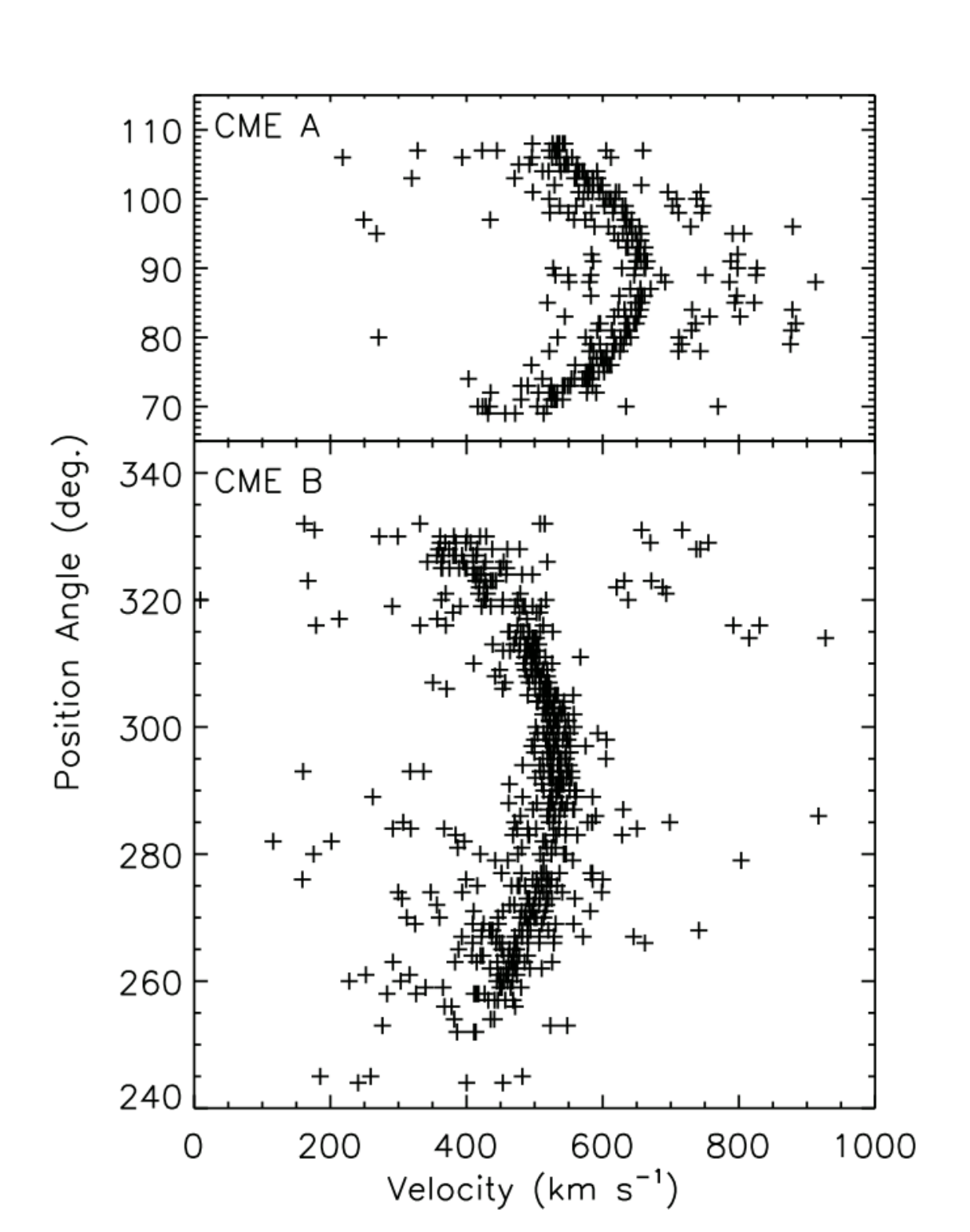}}
\caption{The derived velocities of CMEs A (top) and B (bottom) for each position angle of the corresponding detections displayed in Figure~\ref{figure_model_stack}. The velocities are shown to cluster in such a manner as to indicate an appreciable expansion of each CME, with the flanks moving slower than the apex in both cases. This is an important characteristic when considering the forces acting on a CME as it propagates.}
\label{figure_model_vel_angles_CMEsAB}
\end{figure}

This methodology has the benefit of obtaining height measurements across the complete span of angles along which the CME propagates. This results in a spread of height-time profiles that represents the different speeds attained along the expanding CME. This is an important property when considering the forces that affect CME propagation and expansion, especially when compared to observations further out in the corona with the Heliospheric Imagers \citep[HI;][]{2009SoPh..254..387E}, or Solar Mass Ejection Imager \citep[SMEI;][]{2004SoPh..225..177J} for example, or indeed compared to in-situ measurements as it evolves into an interplanetary CME. Figure~\ref{figure_model_vel_angles_CMEsAB} demonstrates this capability for the relatively large flux-rope CMEs A and B. This will allow a min, max, mean and/or median etc. velocity and acceleration to be determined, along with any changes to the position, trajectory, and angular width of the event.

For the purposes of illustrating the automated detection technique, the above models were propagated with constant velocities of 600\,km\,s$^{-1}$ for CMEs A\,--\,C and 500\,km\,s$^{-1}$ for CMEs D\,--\,F (a model with non-constant acceleration is also discussed below). Their apparent speeds are different due to the different longitudinal directions of propagation (see Table~1 of Paper~\RNum{1}). In order to retrieve the velocities of the model CMEs, the detection stack is inspected as follows:
\begin{enumerate}
\item The detection regions are cleaned and grouped as discussed in Section~\ref{sect_kins}.
\item The height measurements along each position angle occurring in a given detection region are recorded.
\item The velocity distribution is derived using a 3-point Lagrangian interpolation on the resultant height-time data set.
\end{enumerate}
Note at this point that the algorithm does not fully distinguish the height-time profile of CME E from CME D, but rather determines it to be trailing material since it is detected in such close proximity behind CME D. This highlights the current limitation the automated methods have in separating the height measurements of co-temporal, co-spatial CMEs.

\begin{figure*}[!p]
\centerline{\includegraphics[scale=0.85, clip=true, trim=0 0 0 0]{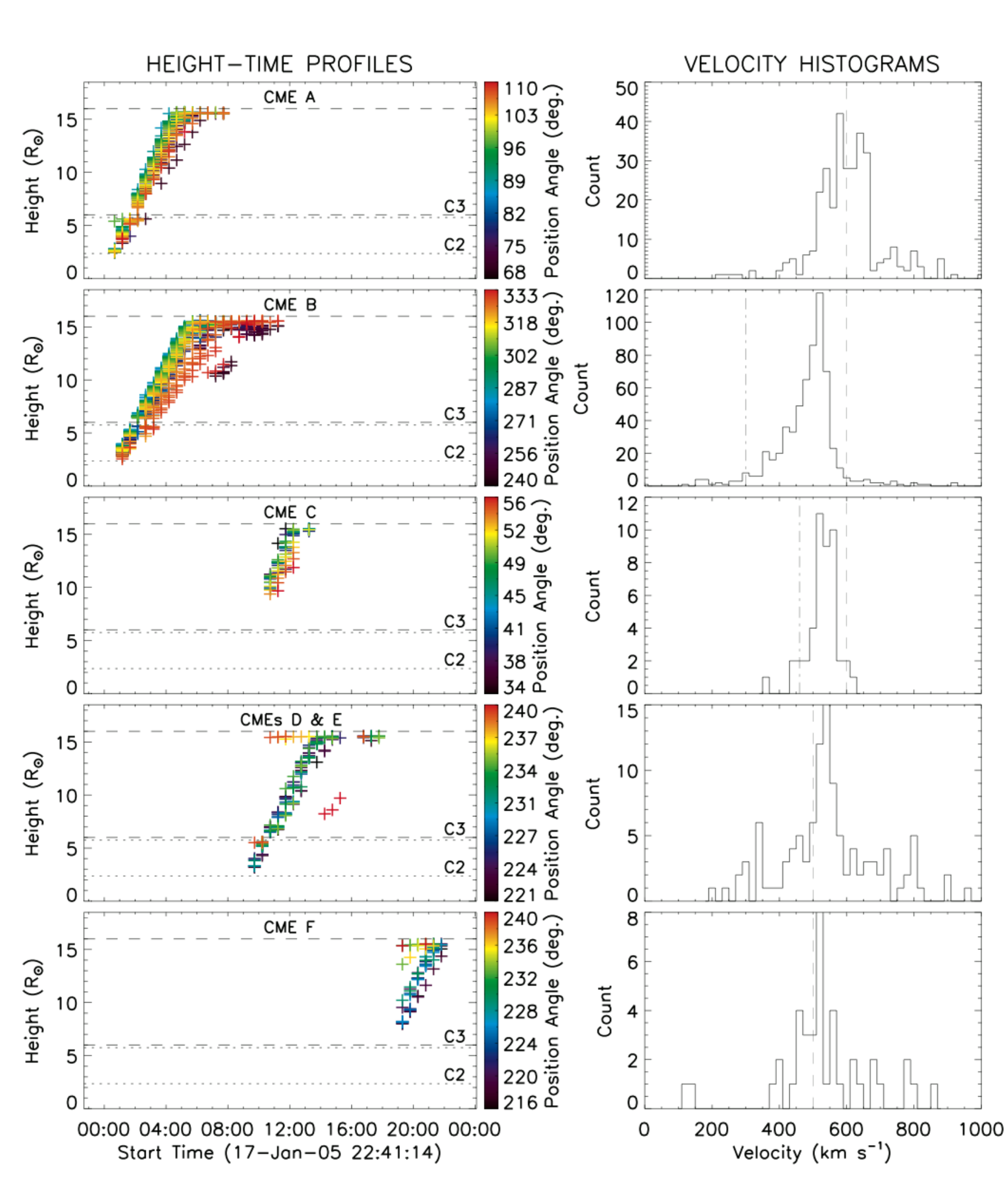}}
\caption{Left: The height-time profiles resulting from the regions in the model detection stack (Fig.~\ref{figure_model_stack}) corresponding to CMEs A through F, where the detection of CME E is not distinguishable from CME D. The position angle corresponding to each height measurement is indicated by the associated colorbar. Right: The derived velocities, displayed in bins of 20~km~s$^{-1}$, where the dashed lines indicate the model velocities of 500 and 600~km~s$^{-1}$, and dot-dashed lines for CMEs B and C indicate the limit of apparent velocity deviation due to projection effects (300 and 460~km~s$^{-1}$ respectively).}
\label{figure_model_histograms}
\end{figure*}

Figure~\ref{figure_model_histograms} shows the resulting height-time measurements for the detection regions corresponding to CMEs A, B, C, D\,\&\,E, and F, and histograms of their corresponding velocity distributions in bins of 20~km~s$^{-1}$. A correction, via a simple histogram segmentation, has been put in place on the velocity output here to ignore the trailing material detections that cause a cluster of zero velocity measurements in the results (i.e., to ignore the height-time measurements corresponding to trailing material once the CME front has left the field-of-view). Thus the histograms of velocity measurements may be deemed to correspond only to the propagating model CME fronts. The input model velocities of 500 and 600~km~s$^{-1}$ are indicated by the dashed lines, while the dot-dashed lines indicate the limit of apparent velocity deviation due to projection effects, which skew the measured velocities of CMEs B and C towards a limit of 300 and 460~km~s$^{-1}$ respectively. A 1\,$\sigma$ interval on the peak of each of the velocity distributions overlaps the known model velocity, even for the CMEs suffering projection effects. Thus, it is deemed that the automated detection and tracking techniques satisfactorily determine the correct height-time profiles of the models, thereby verifying their applicability and robustness.

\begin{figure}[!t]
\centerline{\includegraphics[scale=0.39]{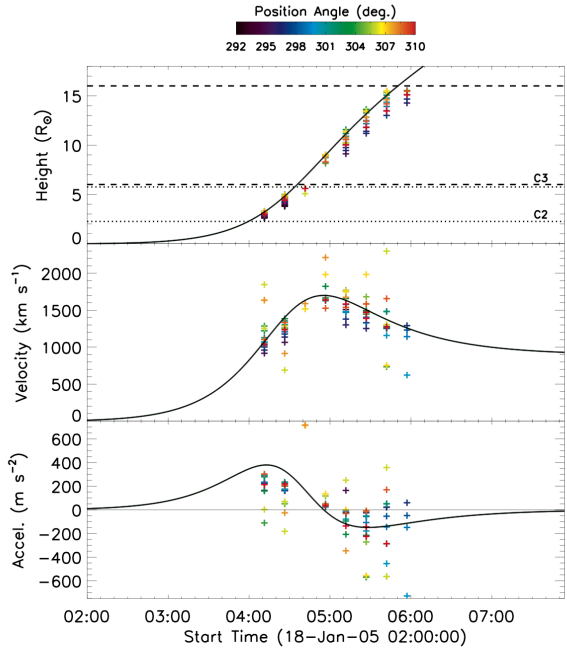}}
\caption{A non-constant acceleration profile input to a model flux-rope CME, and the resulting derived kinematics from the CORIMP automated detection algorithms. The solid curves are the model kinematics, and the `plus' symbols are the resulting kinematics from the automated detection algorithms with a colorbar indicating their relevant position angles (measured counter-clockwise from solar north).}
\label{figure_peak_accel}
\end{figure}

Another model CME flux-rope was generated to test how the automated methods would fare with regards to deriving a non-constant acceleration profile, specifically one which exhibits an initial peak followed by a deceleration and then leveling to zero. This is akin to a general impulsive CME that undergoes an initial high acceleration and then decelerates to match the solar wind speed. The model kinematic profiles are described by the following equations, based on a variation of the acceleration function chosen by \citet{2003ApJ...588L..53G}:

\begin{eqnarray}
h(t)\,=&\,\sqrt{2x}\,t\tan^{-1}\left(\frac{e^{t/2x}}{\sqrt{2x}}\right) \\
v(t)\,=&\,\sqrt{2x}\tan^{-1}\left(\frac{e^{t/2x}}{\sqrt{2x}}\right)+\frac{e^{t/2x}t}{e^{t/x}+2x} \\
a(t)\,=&\,\frac{e^{t/2x}\left(2x\left(t+4x\right)-e^{t/x}\left(t-4x\right)\right)}{2x\left(e^{t/x}+2x\right)^2}
\end{eqnarray}

where $x$ is a scaling factor, set at $x$\,=\,1200 for this case. Figure~\ref{figure_peak_accel} shows the model CME kinematics (solid line) and the over-plotted height, velocity, and acceleration measurements resulting from the CORIMP automated detection and tracking of the CME. The 3-point Lagrangian interpolation is prone to some scatter, especially at the end-points which are therefore less reliable. The kinematic trends of the model CME are, however, satisfactorily revealed by the methods. It is clear that the limits of the observations (restricted fields-of-view, cadence, measurement errors) can dramatically affect their derivation. For example, there are only two satisfactory measurements in C2 before the majority of the CME front leaves the field-of-view, and similarly for the final measurement in C3 where some of the CME front has already left the field-of-view. Nonetheless, given these inherent limitations of the data, the automated methods still prove accurate and effective. 

Ongoing efforts in this vein will lead to a catalogue of real data that can list the determined velocity and acceleration of a CME, as well as the aforementioned parameters of CME height, observation time, position angle, trajectory, and angular width, plus the detailed edge detections outlining the inherent structure of the ejected material of each event. It is intended to utilize this method of cataloguing for integrating CME detections into the Heliophysics Event Knowledgebase (HEK) through collaboration with the Solar Dynamics Observatory Feature Finding Team \citep[SDO FFT;][]{2012SoPh..275...79M}.

\section{Conclusions}
\label{sect_conclusions}

The main objective in implementing an automated detection and tracking routine is to output reproducible, robust, accurate CME measurements (height, width, position angle, etc.). Current methods of CME detection have their limitations, mostly since these diffuse objects have been difficult to identify using traditional image processing techniques. These difficulties arise from the transient nature of the CME morphology, the scattering effects and non-linear intensity profile of the surrounding corona, the presence of coronal streamers, and the addition of noise due to cosmic rays and solar energetic particles (SEPs) that impact the coronagraph detector, along with instrumental effects of stray light, the limitations imposed by low cadence observations, and data corruption or dropouts. In the introduction to this paper, the drawbacks of current cataloguing procedures for investigating CME dynamics (CDAW, CACTus, SEEDS, ARTEMIS) were highlighted as the motivation for establishing a new catalogue. However, given the highly variable nature of CME phenomena and the coronal atmosphere they traverse, there are certain limitations that can never be overcome but only minimized; and it is exactly such a minimizing of current limitations that these new CORIMP methods achieve. The methods are completely automated, making them robust and reproducible - important for back-dating the full LASCO dataset and inspecting the statistics across thousands of events. The automated detection has been extended through both the LASCO/C2 and C3 fields-of-view without any need for differencing, thus minimizing the issues of under-sampling events and of the uncertainty involved when subtracting and scaling images. The multiscale filtering technique reveals the CME structure and so minimizes the uncertainty in determining their often complex geometry. The number of scales in the multiscale decomposition also allows a strength of detection to be assigned through both the magnitude and angular information, thus minimizing the chances that a CME, or parts thereof, go undetected. Furthermore, the spread of measurements available for inspection of the CME kinematics minimizes the uncertainty involved when deriving velocity and acceleration profiles, which is important for comparing with physical theory of CME propagation. Indeed, the overall CORIMP method of automatically detecting, tracking, and deriving CME parameters has been described and demonstrated here on a number of well-conceived models, and real data, with excellent results.



\acknowledgments

This work is supported by SHINE grant 0962716 and NASA grant NNX08AJ07G to the Institute for Astronomy. We thank the anonymous referee for their valuable comments. The SOHO/LASCO data used here are produced by a consortium of the Naval Research Laboratory (USA), Max-Planck-Institut fuer Aeronomie (Germany), Laboratoire d'Astronomie (France), and the University of Birmingham (UK). SOHO is a project of international cooperation between ESA and NASA. The STEREO/SECCHI project is an international consortium of the Naval Research Laboratory (USA), Lockheed Martin Solar and Astrophysics Lab (USA), NASA Goddard Space Flight Center (USA), Rutherford Appleton Laboratory (UK), University of Birmingham (UK), Max-Planck-Institut f\"{u}r Sonnen-systemforschung (Germany), Centre Spatial de Liege (Belgium), Institut d'Optique Th\'{e}orique et Appliqu\'{e}e (France), and Institut d'Astrophysique Spatiale (France).

\bibliographystyle{apj.bst}
\bibliography{references.bib}

\end{document}